\documentclass[12pt]{article}

\def\hybrid{\topmargin -20pt  \oddsidemargin 0pt
       \headheight 0pt   \headsep 0pt
       \textwidth 6.25in 
       \textheight 9.5in 
       \marginparwidth .875in
       \parskip 5pt plus 1pt   \jot = 1.5ex}

\hybrid

\def\o+{\oplus}

\def\beqa{\begin{eqnarray}}
\def\eeqa{\end{eqnarray}}
\def\del{\partial}

\newcommand{\be}[3]{\begin{equation}  \label{#1#2#3}}
\newcommand{\bea}[3]{\begin{eqnarray}  \label{#1#2#3}}
\newcommand{\ee}{\end{equation}}
\newcommand{\eea}{\end{eqnarray}}
\newcommand{\ba}{\begin{array}}
\newcommand{\ea}{\end{array}}
\newcommand{\ov}{\overline}

\begin{document}

\thispagestyle{empty}
\rightline{LMU-ASC 61/05}
\rightline{MPP-2005-94}
\rightline{hep-th/0509082}
\vspace{1truecm}
\centerline{\bf \LARGE Fermion Zero Modes in the Presence of Fluxes}
\vspace{0.3truecm}
\centerline{\bf \LARGE and a Non-perturbative Superpotential}

\vspace{1truecm}
\centerline{}

\vspace{1truecm}
\centerline{Dieter L\"ust$^{a,b,}$\footnote{luest@mppmu.mpg.de, luest@theorie.physik.uni-muenchen.de}, Susanne
Reffert$^{b,}$\footnote{sreffert@theorie.physik.uni-muenchen.de},
Waldemar Schulgin$^{b,}$\footnote{schulgin@theorie.physik.uni-muenchen.de}
and Prasanta K. Tripathy$^{a,}$\footnote{prasanta@theorie.physik.uni-muenchen.de}}

\vspace{.7truecm}

\centerline{{\em $^a$Arnold-Sommerfeld-Center for Theoretical Physics}}
\centerline{{\em Department f\"ur Physik,
Ludwig-Maximilians-Universit\"at  M\"unchen}}
\centerline{{\em Theresienstra\ss e 37, 80333 M\"unchen, Germany}}

\vspace{.4truecm}

{\em \centerline{$^b$Max-Planck-Institut f\"ur Physik,
F\"ohringer Ring 6, 80805 M\"unchen, Germany}}

\vspace{.4truecm}

\vspace{.7truecm}


\vspace{.4truecm}



\begin{abstract}\noindent
We study the effect of background fluxes of general Hodge type
on the supersymmetry conditions and on
the fermionic zero modes on the world-volume of a 
Euclidean $M5/D3$--brane in $M$-theory/type $IIB$ string theory.

\vskip0.2cm
\noindent
Using the naive susy variation of the modulino fields to 
determine the number of zero modes in the presence of a flux of general Hodge type, an 
inconsistency appears. This inconsistency is 
resolved by a modification of the supersymmetry variation of the 
modulinos, which captures the back-reaction of the non-perturbative effects on the background flux and the geometry.
\end{abstract}

\newpage
\pagenumbering{arabic}

\section {Introduction and Summary}

Recently, there has been a lot of progress in the 
investigation of KKLT--type models \cite{hep-th/0301240}. 
On the one hand, specific examples of candidate models 
have been constructed \cite{hep-th/0404257, hep-th/0503124}. 
On the other hand, the generation of a non-perturbative superpotential which 
may serve to stabilize all K\"ahler moduli 
has been investigated in much detail. 
The recent research in this line 
extends the earlier work of Witten \cite{hep-th/9604030} by 
taking into account non--vanishing background 
fluxes \cite{hep-th/0501081, hep-th/0503072, hep-th/0503125, hep-th/0503138, hep-th/0504041, hep-th/0504058} 
and working out the conditions for the generation of the superpotential directly for type $IIB$--orientifolds 
without the detour of analyzing 
the $M/F$--theory case first \cite{hep-th/0507069, hep-th/0507091, hep-th/0506179}. 
If $M5/D3$--brane instantons wrapping a divisor in the compactification 
manifold are the source of a possible non-perturbative superpotential,  the analysis involves 
deriving the Dirac equation in the world--volume of 
the $M5/D3$--brane and studying the structure of its fermionic zero modes.
So far, only the case of the background flux being of 
Hodge type $(2,2)$ in $M/F$--theory, or $(2,1)$ in type $IIB$--theory has been considered.

The present letter resolves a seeming puzzle concerning the 
fermionic zero mode structure in the presence of background fluxes of general Hodge type.
As has been 
shown in \cite{hep-th/0502168, hep-th/0506090}, 
the conditions for a supersymmetric background flux obtained from the minimization 
of the effective four-dimensional superpotential change in the presence of a non-perturbative term. 
The supersymmetric flux is no longer of Hodge type $(2,2)$ (resp. $(2,1)$ for type $IIB$), 
but receives contributions of all Hodge types. We will show that,
if one now, guided by this result, 
plugs a flux of general Hodge type into the 
zero mode conditions obtained from the Dirac equation, 
an inconsistency arises: If with $(2,2)$--flux, the conditions for 
the generation of a superpotential were met, this is no longer the case for general flux. 

As we explain in the following, this apparent mismatch disappears 
after the introduction of a modification of the supersymmetry variation of the modulino, 
which basically captures the back-reaction of the 
non-perturbative effects on the background flux and the geometry.

This paper is organized as follows. 
In section 2, we discuss the supersymmetry conditions 
on the background flux as obtained from the effective 
potential as well as from the supersymmetry variation of the modulino.
In section 3,  the fermionic zero modes on the world--volume of $D3/M5$--branes are analyzed. 
First, we study the conditions on the zero modes originating from the Dirac 
equation in the presence of fluxes of general Hodge type. We find a modification from the 
case of pure $(2,2)$, (respectively $(2,1)$) flux. 
To further elucidate the effect of allowing general background fluxes, 
the concrete example of compactification of $M/F$--theory on $K3\times K3$ is presented. 
If a pure $(2,2)$--flux is turned on, a non-perturbative superpotential is generated. 
If, on the other hand, we allow other Hodge components, which a supersymmetric 
flux solution in the presence of a non-perturbative superpotential requires, 
all zero modes are lifted and no non-perturbative superpotential is generated. 
Thus, an obvious inconsistency arises.

In section 4, we set out to resolve the puzzle. 
We find that the non-perturbative superpotential must be 
included into the susy variation of the 
11--dimensional gravitino field after compactification, 
which in turn determines the Dirac equation and therefore the number of zero modes. 
Like this, the $(4,0)$-- and $(3,1)$--parts of the flux are balanced by the 
contribution from the non--perturbative superpotential, and the number of 
zero modes remains the same as for $(2,2)$--flux.

\section{The supersymmetry conditions}

In this section, we will study the supersymmetry conditions for the low energy
theory arising from string compactifications in the presence of background fluxes 
and due to non-peturbative effects. We will first obtain the susy conditions 
by minimizing the effective potential and then by analyzing the spinor 
conditions. We work out the case for $IIB$ compactification on Calabi-Yau
threefolds and then briefly discuss $M$--theory compactification on 
Calabi-Yau fourfolds.

\subsection{Effective Potential}

We first consider the compactification of type $IIB$ theory on a Calabi-Yau
threefold. The resulting low energy supergravity action is given by
\begin{eqnarray}
S = \int d^4 x {1\over 2} \sqrt{-g}\left\{
R + g_{A\overline B} \del_\mu z^A \del^\mu \overline z^B \right\} 
+ V_{eff} + S_{gauge} \ .
\end{eqnarray}
Here, we used a condensed notation: The indices $\{A,B,\cdots\} = \{i,I,\tau\}$ 
denote both the complex structure moduli $\{i\}$, K\"ahler moduli $\{I\}$,
and the complexified axion-dilaton field~$\tau$. $S_{gauge}$ denotes the gauge 
field dependent part of the action. The effective potential 
\beqa
V_{eff} = {1\over 2} e^K \left( g^{AB} D_AW \overline{D_BW} - 3 |W|^2\right)
\eeqa
is given in terms of the total superpotential 
\beqa
W = W_{flux} + W_{np}
\eeqa
and the  K\"ahler potential $K$. Here $W_{flux}$ is the flux superpotential
\cite{hep-th/9912152}
\beqa
W_{flux} =  \int G_3 \wedge \Omega_3 \ ,
\eeqa
and $W_{np}$ is the superpotential arising from nonperturbative effects.
$\Omega_3$ is the holomorphic $(3,0)$--form on the CY space and 
\beqa
G_3 = F_3 - \tau H_3 ~,
\eeqa
 $F_3$ and $H_3$ being the RR and NS field strengths, respectively. The flux 
superpotential depends only on the complex structure moduli. We assume the 
nonperturbative superpotential to depend on the K\"ahler moduli only.

The supersymmetry preserving minima are obtained by solving the equations
\beqa 
D_A W = 0 \ .
\label{susy} 
\eeqa
It is well known that in the absence of a nonperturbative term, $W = W_{flux}$, 
the condition (\ref{susy}) requires $G_3$ to be of type $(2,1)$ and primitive 
\cite{hep-th/0105097}. For $W_{np} \neq 0$, this is no longer true 
\cite{hep-th/0506090}, and $G_3$ acquires non-vanishing $(1,2),(3,0)$ and $(0,3)$
parts: 
\beqa
&& \int G_3\wedge \chi_i^{(2,1)} 
+ \del_i K W_{np} = 0 \ ,\cr && 
\int G_3\wedge \Omega_3~\del_I K + D_I W_{np} = 0 \ , \cr
&& \int \overline G_3 \wedge \Omega_3 + W_{np} = 0 \ .
\label{iibsusy}
\eeqa
The primitivity condition $G_3\wedge J = 0$, being a $D$--term condition, remains intact
despite $W_{np}$. Here $\chi_i^{(2,1)}$ is a form of type $(2,1)$.

We can similarly obtain the susy conditions for M-theory compactification on
a Calabi-Yau fourfold. The flux superpotential is now given by
\cite{hep-th/9906070}
\beqa
W_{flux} = \int G_4 \wedge \Omega_4 \ .
\eeqa
Here, $G_4$ is the four-form flux present in 11-dim. supergravity theory and 
$\Omega_4$ is the holomorphic $(4,0)$--form on the CY fourfold.
The susy conditions take the form:
\beqa
&& \int G_4 \wedge \chi^{(3,1)}_i + \del_i K W_{np} = 0\ , \cr
&& \int G_4 \wedge \Omega_4~\del_I K + D_I W_{np} = 0 \ .
\label{mthsusy}
\eeqa
In the following subsection, we will show how the above conditions can be 
derived from the modulino variations.

\subsection{Spinor Conditions}

Now, it is important to remember that the BPS susy variation of the gravitino 
is equivalent to solving the susy conditions in the effective field theory,
as discussed in \cite{hep-th/9605053} for M-theory on a fourfold, in 
\cite{hep-th/0201028} and in \cite{hep-th/0106014} for type $IIB$ on a CY 
threefold, and also by \cite{hep-th/0306088} for the heterotic string. 
Thus we must modify the spinor conditions accordingly in order to obtain
the susy conditions eq.(\ref{iibsusy}) in $IIB$ theory and eq.(\ref{mthsusy})
in M-theory. In what follows, we will first review the spinor conditions 
in the absence of $W_{np}$, and then consider the generalization when $W_{np}$
is included. 

Let us first consider the situation in $IIB$ theory. This has been worked out 
in \cite{hep-th/0106014}. The supersymmetry variations can be 
summarized as follows:
\beqa
\kappa \delta\psi_\mu &=& \partial_\mu \epsilon - {1\over 8} \gamma_\mu 
\gamma^m \left(\del_m \ln Z - 4 \kappa Z \Gamma^4 \del_m h\right)\epsilon 
+ {1\over 16} \kappa \gamma_\mu G \epsilon^* ,\cr 
\kappa \delta\psi_m &=& \left(\tilde D_m - {i\over 2} Q_m\right)\epsilon
+ {1\over 8} \epsilon \del_m \ln Z -  {1\over 16} \kappa \gamma_m G \epsilon^* 
- {1\over 8}\kappa  G \gamma_m \epsilon^* ,\cr 
\kappa \delta\lambda^* &=& - i \gamma^m P_m^*\epsilon + {i\over 4} \kappa
\overline{G} \epsilon^* .
\label{gpsusy}
\eeqa
The first equation is the supersymmetry variation of the four-dimensional
gravitino field. Second, $\delta \psi_m$ corresponds to the variation of
the internal gravitino. After compactification the internal gravitino
degrees of freedom become in the effective 4D field theory the modulino fields,
i.e. the fermionic superpartners of the K\"ahler and complex structure
moduli fields. Concretely, the modulino equations which one obtains by dimesional reduction (see appendix) are
\begin{eqnarray}\label{modulino}
\delta \phi^i_{e\overline{ab}}&=&-\frac{1}{8}G_{e\overline{ab}}^i\hat\xi^*-\frac{1}{16}g^{a\overline c}g_{e\overline a}G^i_{a\overline{bc}}\hat\xi^* \ , \qquad i=1,\ldots ,h^{(2,1)} \ , \cr
\delta\phi^I_{\overline e |\overline{ab}}&=&-\frac{1}{16}G_{\overline{eab}}\hat\xi^* \ , \ \ \ \qquad\qquad\qquad\qquad I=1,\ldots , h^{(1,1)} \ , \cr
\delta\lambda^*_{\overline{abc}}&=&\frac{i}{4}\overline G_{\overline{abc}}\hat\xi^* \ ,
\end{eqnarray}
where $\hat \xi$ is a four dimensional supersymmetry parameter.
Finally, $\delta \psi_m$ indeed
comprises the supersymmetry variations of all modulinos, namely it
leads after compactification to $h_{1,1}+h_{2,1}$ independent spinor equations, which we call
modulino equations. Finally, $\delta\lambda^*$ is the supersymmetry variation
of the four-dimensional dilatino.   
In these equations, we use the same notation as  \cite{hep-th/0106014}. 
In particular,
$ G = {1\over 6} G_{mnp}\gamma^{mnp}$, $Z$ is the warp factor, $\tilde D_m$
is the covariant derivative with respect to the internal metric, $h$ is 
related to the RR four--form field, $h = C_{0123}$, and 
\beqa
&& P_m = f^2 \del_m B~,~~ Q_m = f^2 {\rm Im} \left(B\del_mB^*\right) \ , \cr
&& B = {1 + i \tau \over 1 - i \tau} ~,~~ f^{-2} = 1 - B B^* .
\eeqa
The conditions (\ref{gpsusy}) can be solved to show that $G_3$ is of 
type $(2,1)$ and primitive.

Clearly, the explicit dependence on the superpotential $W_{flux}$ and 
its covariant derivatives is not apparent in the modulino variations 
(\ref{modulino}). We need to make this precise, in order to generalize the 
above formulae in presence of $W_{np}$.  Since we are interested in the 
$G_3$ dependence of the variations, we can as well ignore the effects of  
warping and the five--form flux, and also set the complexified axion-dilaton
field to constant. 

It is now easy to introduce the flux superpotential in the above equations.
Note that 
\beqa
&& D_i W_{flux} =  \int G_3 \wedge \chi_i^{2,1} 
\Longrightarrow G^i_{a\overline{bc}} = \epsilon_{a\overline{bc}} D_iW_{flux}\ , \cr
&& D_I W_{flux} =  \del_I K \int G_3\wedge\Omega_3
\Longrightarrow G_{\overline{abc}} = \epsilon_{\overline{abc}} { D_I W_{flux}
\over \del_I K}\ .
\eeqa
Substituting the above into the modulino variations, we find
\begin{eqnarray}
\delta\phi_{e\overline {ab}}^i &=& -\frac{1}{8} \epsilon_{e\overline {ab}} D_i W_{flux}
-\frac{1}{16}g^{a\overline c}g_{e\ov a}G^i_{a\ov{bc}}\hat\xi \ , \qquad  i=1\ldots h^{(2,1)} \ , \cr
\delta\phi^I_{\ov e|\ov {ab}} &=& -\frac{1}{16}\epsilon_{\ov{eab}}\frac{D_IW_{flux}}{\partial _I K} \ , \qquad\qquad\qquad\qquad \ \  I=1\ldots h^{(1,1)} \ . 
\end{eqnarray}
Similarly, using 
\beqa
\overline{G}_{\overline{abc}} 
= - \epsilon_{\overline{abc}} (\tau-\overline\tau) D_\tau W_{flux} \ ,
\eeqa
we find
\beqa
\delta\lambda_{\ov {abc}}^* = -{i\over 4} \epsilon_{\ov{abc}} (\tau-\overline\tau) D_\tau W_{flux} 
\hat \xi\ .
\eeqa
For covariantly constant spinors, we recover the susy conditions 
\beqa
D_i W_{flux} = D_I W_{flux} = D_\tau W_{flux} = 0\ .
\eeqa
Now, it is easy to generalize the spinor variations in presence of the 
non-perturbative superpotential. We simply replace 
$W_{flux}$ by $W = W_{flux} + W_{np}$. The variation equations then become
\beqa
\delta\phi_{e\overline {ab}}^i &=& -\frac{1}{8} \epsilon_{e\overline {bc}} D_i W
-\frac{1}{16}g^{a\overline c}g_{e\ov a}G^i_{a\ov{bc}}\hat\xi \ , \cr
\delta\phi^I_{\ov e|\ov {ab}} &=& -\frac{1}{16}\epsilon_{\ov{eab}}\frac{D_IW}{\partial _I K} \ , \cr 
\delta\lambda_{\ov {abc}}^* &=& -{i\over 4} \epsilon_{\ov{abc}} (\tau-\overline\tau) D_\tau W \ .
\eeqa
We clearly see that, for covariantly constant spinors, the first of the above 
equations implies the flux to be primitive and in addition $D_i W$ is zero.
The second and third equations then imply that $D_I W$ and $D_\tau W$ are 
zero respectively. Thus we recover the susy conditions 
\begin{equation}
D_i W = D_I W = D_\tau W = 0 ~.
\end{equation}

We now proceed to work out the modulino transformations in M-theory in 
presence of $W_{np}$ in a similar fashion. This has been analyzed in 
\cite{hep-th/9605053}. We will first express the variation equations 
in terms of the flux superpotential, and then generalize it to the case of
$W_{np}\ne 0$. Consider first the internal gravitino 
variation without $W_{np}$:
\beqa
\delta \psi_m = \nabla_m \xi + {1\over 24} \gamma^{npq} G_{mnpq} \xi \ .
\eeqa
By dimensional reduction we obtain (see appendix)
\begin{eqnarray}\label{holomorph}
\delta\phi_{e\overline c}^k&=&\frac{1}{4}\left(G_{eb\overline{cd}}g^{b\overline d}\right)^k\hat\xi\ , \ \ \ \ \qquad\qquad k=1,\ldots, h^{(1,1)}  \ ,\cr
\delta\phi^i_{e\overline{abc}}&=&\frac{1}{24}G^I_{e\overline{abc}}\hat\xi \ , \qquad\qquad\qquad\qquad i=1,\ldots, h^{(3,1)} \ , \cr
\delta\phi^I_{\overline e | \overline{abc}}&=&\frac{1}{24}G_{\overline{eabc}}\hat\xi\ , \qquad\qquad\qquad\qquad I=1,\ldots, h^{(1,1)} \ .
\end{eqnarray}
By solving the susy conditions,  we get in general 
$h_{3,1}$ equations for the complex structure moduli and $h_{1,1}$ equations 
for the K\"ahler moduli. 
The same conditions should be reproduced by setting 
$\delta \phi^i$ and $\delta \phi^I$ to zero.
There are $h^{3,1}$ fluxes of type $(1,3)$. 
The $(0,4)$-flux is 
a solution of $h^{1,1}$ independent equations. Because of these reasons,  it is 
natural to say that for every $G_{a\overline{bcd}}$ and every $G_{\overline{abcd}}$
(same $G_{\overline{abcd}}$ coming from $h^{1,1}$ equations), the variation of the gravitino should be zero.

There is no $I$ on the r.h.s. This emphasizes the fact that the $h^{1,1}$ supersymmetry 
conditions are 
degenerate in the $(0,4)$-flux.
Using 
\begin{eqnarray}\label{Wflux1}
D_i W_{flux} = \int G_4 \wedge \chi^i_{ 3,1} = G^i_{ e \overline{bcd}} 
\epsilon^{e\overline{bcd}}
\end{eqnarray}
and
\begin{eqnarray}\label{Wflux2}
D_I W_{flux} = \partial_I K \int G_4\wedge \Omega_4 =
\partial_I K G_{\overline {ebcd}} \epsilon^{\overline{ebcd}} \ ,
\end{eqnarray}
we can immediately rewrite (\ref{holomorph}) into
\begin{eqnarray}\label{eqdw}
\delta\phi_{e\overline c}^k&=&\frac{1}{4}\left(G_{eb\overline{cd}}g^{b\overline d}\right)^k\hat\xi\ , \ \ \ \ \ \ \ \ \qquad k=1,\ldots, h^{(1,1)}  \ ,\cr
\delta\phi^i_{e\overline{abc}}&=&\frac{1}{24}\epsilon_{e\ov{abc}}D_iW_{flux}\hat\xi \ , \qquad\qquad i=1,\ldots, h^{(3,1)} \ , \cr
\delta\phi^I_{\overline e | \overline{abc}}&=&\frac{1}{24}\epsilon_{\overline{eabc}}\frac{D_I W_{flux}}{\partial_ I K}\hat\xi\ , \qquad\qquad I=1,\ldots, h^{(1,1)} \ .
\end{eqnarray}

The supersymmetry conditions and the primitivity condition are reproduced by setting 
$\delta \phi ^k$, $\delta {\phi^i}$, $\delta {\phi^I}$ to zero.

This gives immediately
\begin{eqnarray}
&& g^{a\bar d}g^{b\bar c}G_{eb\overline{cd}}=0 \ , \cr
&& D_i W_{flux} = 0 \ ,\ i=1,\ldots ,h^{1,3} \ , \cr
&& D_I W_{flux} = 0 \ ,\ I=1,\ldots , h^{1,1} \ .
\end{eqnarray}
These equations correspond to the primitivity conditions on $G_{2,2}$ 
and the vanishing of $G_{1,3}$ and $G_{0,4}$.

In the next step, we would like to make a proposal for the form 
of the additional terms of the supersymmetry variation of the modulinos 
in the presence of the non-perturbative term $W_{np}$. 
The supersymmetry conditions which should be reproduced, change to
\begin{eqnarray}
&& D_i W = D_iW_{flux}+D_i W_{np}=0 \ ,\cr
&& D_I W = D_IW_{flux}+D_I W_{np}=0 \ . 
\end{eqnarray}
From ($\ref{eqdw}$), we immediately see that the variation of the modulinos should be changed to
\begin{eqnarray}\label{modulinomod}
\delta\phi_{e\overline c}^k&=&\frac{1}{4}\left(G_{eb\overline{cd}}g^{b\overline d}\right)^k\hat\xi\ , \ \ \ \ \ \ \ \ \qquad k=1,\ldots, h^{(1,1)}  \ ,\cr
\delta\phi^i_{e\overline{abc}}&=&\frac{1}{24}\epsilon_{e\ov{abc}}D_iW\hat\xi \ , \qquad\qquad i=1,\ldots, h^{(3,1)} \ , \cr
\delta\phi^I_{\overline e | \overline{abc}}&=&\frac{1}{24}\epsilon_{\overline{eabc}}\frac{D_I W}{\partial_ I K}\hat\xi\ , \qquad\qquad I=1,\ldots, h^{(1,1)} \ .
\end{eqnarray}


\section{Conditions on the zero modes from fluxes and the non-perturbative
superpotential}

The non-perturbative superpotential may be generated via gaugino 
condensation or via instanton effects or both. 
Here, we will concentrate on the case of instantons. In type $IIB$ theory, they correspond to
Euclidean $D3$--branes wrapping divisors of the CY threefold, whereas in 
M-theory, they come from Euclidean $M5$--branes wrapping divisors of the CY 
fourfold. It has been pointed out by Witten \cite{hep-th/9604030} some 
time ago that the necessary condition for an $M5$--instanton to generate a 
superpotential is that the corresponding divisor has holomorphic Euler 
characteristic equal to one. This provides a stringent condition on the possible 
CY fourfolds \cite{hep-th/0405011}. For type $IIB$ compactification on a Calabi-Yau without the orientifold projection
(without flux),  the index is always zero and hence 
no superpotential is generated due to instanton effects \cite{hep-th/0507069}.
It has been argued recently \cite{hep-th/0407130}, that the index might change 
in the presence of flux. An explicit example has been constructed to 
show that some of the wold--volume fermion zero modes are lifted due to flux
\cite{hep-th/0503072}. Subsequently, a generalized index formula was derived 
in M-theory \cite{hep-th/0503125,hep-th/0503138}, as well as in type $IIB$ 
theory \cite{hep-th/0507069}. However, these results are based on the assumption
that the flux is primitive and of type $(2,1)$ in type $IIB$, or  $(2,2)$ respectively in 
M-theory.  As we have already discussed, the supersymmetric flux no longer remains $(2,1)$ 
(resp. $(2,2)$) in presence of the non-perturbative superpotential. In this section, 
we will analyze the fermion zero modes on the world volume of $D3$/$M5$--branes  
in the presence of general flux.

\subsection{General fluxes}

The fermionic bilinear terms in the $D3$--brane world--volume action in presence 
of background flux have been derived in \cite{hep-th/0202118,hep-th/0503072}
by using the method of gauge completion, and also in \cite{hep-th/0303209,
hep-th/0306066,hep-th/0504041} from the $M2$--brane world volume action using T-duality.
Upon Euclidean continuation and by an appropriate gauge choice
\cite{hep-th/0507069}, the Lagrangian takes the form 
\beqa
L^{D3} = 2 \sqrt{{\rm det}~g} ~~ \theta\left\{ 
e^{-\phi} \gamma^m \nabla_m + {1\over 8} \tilde G_{mn\hat p}
\gamma^{mn\hat p}\right\} \theta \ .
\eeqa
Here $m,n,\ldots$ are directions along the brane and $\hat p$ stands for
directions transverse to the brane.  
As always, we turn on the three--form flux 
only along the directions of the internal manifold. Also for simplicity, we 
set the flux $F_2$ due to the world--volume gauge fields to zero. 
$\tilde G$ is defined to be 
\beqa
\tilde G_{mnp} = e^{-\phi} H_{mnp} + i F'_{mnp} \gamma_5 \ ,
\eeqa
with $F' = dC_2 - C_0 H_3$. The Dirac equation, obtained from the above action,
reads
\beqa
\left\{e^{-\phi} \gamma^m \nabla_m 
+ {1\over 8} \tilde G_{mn\hat p}\gamma^{mn\hat p}\right\} \theta = 0 \ .
\eeqa
Locally, we can express the internal metric as 
\beqa
ds^2 = g_{a\bar b} dy^a dy^{\bar b} + g_{z\bar z} dz d\bar z \ ,
\eeqa
where $a,b,\ldots\, $ are complex coordinates on the $D3$--brane and 
$z,\overline z$ are directions transverse to the brane. We define the Clifford
vacuum to be 
\beqa
\gamma^z |\Omega> = \gamma^a |\Omega> = 0\ .
\eeqa
The spinor $\theta$ can be written in terms of positive and negative chirality 
spinors as $\theta = \epsilon_+ + \epsilon_-$ with 
\beqa
\epsilon_+ &=& \phi |\Omega> + \phi_{\overline a} \gamma^{\overline a} |\Omega>
+ \phi_{\overline{ab}} \gamma^{\overline{ab}} |\Omega>\ , \cr
\epsilon_- &=& \phi_{\overline z} \gamma^{\overline z} |\Omega> 
+ \phi_{\overline{az}} \gamma^{\overline{az}} |\Omega>
+ \phi_{\overline{abz}} \gamma^{\overline{abz}} |\Omega> \ .
\eeqa
Substituting this into the Dirac equation, we find 
\beqa
e^{-\phi} 2 g^{a\overline a}\del_a\phi_{\overline a}
+ 2 i g^{z\overline z} g^{a\overline{b'}}g^{b\overline{a'}} G_{abz} 
\phi_{\overline{a'b'z}}
+{1\over 2} i g^{z\overline z}g^{a\overline b} \phi_{\overline z} 
G_{a\overline b z} = 0\ , \cr
e^{-\phi} \left(\partial_{\overline{a'}}\phi
+ 4 g^{a\overline{b'}} \del_a \phi_{\overline{b'a'}}\right) 
+ {1\over 2} i  g^{z\overline z} g^{a\overline b} 
\left(\phi_{\overline{a'z}} \overline{G}_{a\overline b z}
- 2 \phi_{\overline{bz}} \overline{G}_{a\overline a z}\right) = 0\ , \cr
e^{-\phi}\del_{[\overline{a'}} \phi_{\overline{b'}]}
+ {1\over 2} i g^{z\overline z} g^{a\overline b}
\left(\phi_{\overline{a'b'z}} G_{a\overline b z}
- 4 \phi_{\overline{bb'z}} G_{a\overline a' z} \right) 
+ {1\over 4} i g^{z\overline z} \phi_{\overline z} G_{\overline{a'b'}z} = 0 
\eeqa 
and 
\beqa
e^{-\phi} 2 g^{a\overline a}\del_a\phi_{\overline{az}}
+ i g^{a\overline{b'}}g^{b\overline{a'}}
\phi_{\overline{a'b'}} G_{ab\overline z} 
+ {1\over 4} i g^{a\overline b} \phi G_{a\overline{bz}} = 0\ , \cr
e^{-\phi} \left(\partial_{\overline{a'}}\phi_{\overline z}
+ 4 g^{a\overline{b'}} \del_a \phi_{\overline{b'a'z}}\right) 
- {1\over 4} i g^{a\overline{b}} \left(\phi_{\overline{a'}}
\overline{G}_{a\overline{bz}} - 2 \phi_{\overline{b}} 
\overline{G}_{a\overline{a'z}}\right) = 0 \ , \cr
e^{-\phi}\del_{[\overline{a'}} \phi_{\overline{b']z}}
+ {1\over 4}  i g^{a\overline{b}} \left(
\phi_{\overline{a'b'}} G_{a\overline{bz}}
- 4 \phi_{\overline{bb'}} G_{a\overline{a'z}} \right)
+ {1\over 8} i \phi G_{\overline{a'b'z}} = 0\ .
\eeqa

We can similarly work out the equations for world--volume $M5$--brane fermions.
The fermionic bilinear terms on the $M5$--brane world--volume in the presence of 
background flux have been derived in \cite{hep-th/0501081}.  Upon setting the 
world volume gauge flux to zero, we have the Dirac equation
\begin{eqnarray}
\gamma^{m} \nabla_{m} \theta - {1\over 24} \gamma^{\hat q}
\gamma^{mnp} G_{mnp\hat q} \theta = 0\ .
\end{eqnarray}
Again, we turn on the fluxes only along the compact directions.  Here, 
$m,n,p, \ldots\,$ are real indices. A `$\,\hat{}\,$' indicates the directions 
transverse to the brane. We denote by
 $a,b, \ldots\,$ the holomorphic indices along the brane and by $\bar a,\bar b,
\ldots $ the anti--holomorphic indices;  $z$ is the complex coordinate
along the normal to the 
divisor. The spinor $\theta$ can 
be expressed in terms of the Clifford vacuum and the creation operators as 
\begin{eqnarray}
\theta = \phi |\Omega> + \phi_{\bar z} \gamma^{\bar z} |\Omega> 
+ \phi_{\bar a\bar b} \gamma^{\bar a\bar b} |\Omega>
+ \phi_{\bar z\bar a\bar b} \gamma^{\bar z\bar a\bar b} |\Omega>\,.
\label{thetaexp}
\end{eqnarray}

Plugging this expression for $\theta$ into the Dirac equation, we find
\begin{eqnarray}
&& (\partial_{\bar c} \phi + 4 g^{b\bar b'} \partial_b \phi_{\bar b'\bar c}) \cr
&& + {1\over 2} \left[ 4 g^{a\bar a'} g^{b\bar b'} g^{z\bar z}
(G_{ab\bar b'z} \phi_{\bar z\bar a'\bar c} 
- G_{ab\bar cz} \phi_{\bar z\bar a'\bar b'}) + g^{z\bar z} g^{a\bar b} \phi_{\bar z}
G_{a\bar b\bar c z} \right] = 0\ , \cr
&& (\partial_{\bar a} \phi_{\bar z} 
+ 4 g^{b\bar b'} \partial_b \phi_{\bar z\bar a \bar b'} )  \cr
&& - {1\over 4} \left[ 4 g^{a\bar a'} g^{b\bar b'} (G_{ab\bar b'\bar z} 
\phi_{\bar a'\bar c} - G_{ab\bar c\bar z} \phi_{\bar a'\bar b'})
+ g^{a\bar b} \phi G_{a\bar b\bar c\bar z}\right] = 0 \ ,\cr
&& \partial_{[\bar a} \phi_{\bar b\bar c]} 
+ {1\over 12} g^{z\bar z}\phi_{\bar z} G_{\bar a\bar b\bar c z} = 0\ , \cr
&& \partial_{[\bar a} \phi_{\bar z\bar b\bar c]} + {1\over 24} \phi
G_{\bar a\bar b\bar c\bar z} = 0\ .
\end{eqnarray}
These expressions can be simplified a lot using the primitivity condition:
\begin{eqnarray}
&& (\partial_{\bar c} \phi + 4 g^{b\bar b'} \partial_b \phi_{\bar b'\bar c}) 
- 2 g^{a\bar a'} g^{b\bar b'} g^{z\bar z}
 G_{ab\bar cz} \phi_{\bar z\bar a'\bar b'} = 0\ , \cr
&& (\partial_{\bar a} \phi_{\bar z} 
+ 4 g^{b\bar b'} \partial_b \phi_{\bar z\bar a \bar b'} )  
+  g^{a\bar a'} g^{b\bar b'}  G_{ab\bar c\bar z} \phi_{\bar a'\bar b'}
= 0\ , \cr
&& \partial_{[\bar a} \phi_{\bar b\bar c]} 
+ {1\over 12} g^{z\bar z}\phi_{\bar z} G_{\bar a\bar b\bar c z} = 0\ , \cr
&& \partial_{[\bar a} \phi_{\bar z\bar b\bar c]} + {1\over 24} \phi
G_{\bar a\bar b\bar c\bar z} = 0\ .
\end{eqnarray}
The equations are modified due to the $(3,1)$-- and $(4,0)$--fluxes, and
so is the zero mode counting. To understand this better, we shall turn to the 
example of compactification on $K3\times K3$.

\subsection{Example: $K3\times K3$ }

To acquire a better understanding of the above equations, we consider here 
the example of M/F-theory compactified on $K3_1\times K3_2$ with background
flux \cite{hep-th/0501139, hep-th/0504058, hep-th/0506014}. Consider one of the $K3$s 
(say $K3_2$) to be elliptically fibered. Wrap the $M5$--brane on one of the 
divisors of the form $K3\times S$, where $S$ corresponds to the $P^1$s of the 
elliptic $K3$. Let $z$ parametrize the direction normal to the brane.

We will now briefly review the case of the flux being of type $(2,2)$ and 
primitive and then consider the case of general flux. Let us first 
analyze the case of the flux preserving $N=2$ supersymmetry.
In this case, the $(2,2)$--flux must take the form
\begin{equation}
G_4 \in H^{1,1}(K3_1) \otimes H^{1,1}(K3_2) ~,
\label{neq2flux}
\end{equation}
which implies that the $N=2$ flux must be a $(1,1)$--form in $K3_2$. Since it 
is an elliptically fibered $K3$, we have to use the spectral sequence, which tells 
us that the flux belongs to \cite{hep-th/9611137}
$$ H^0(B, R^2\pi_* {\bf R}) \oplus H^2(B,\pi_*{\bf R})\ , $$
which in simple terms means that the flux has either both legs in the
fiber or both in the base. So the $N=2$ flux is always of the type 
$ G_{a\bar b c\bar d} ~{\rm or}~ G_{a\bar b z\bar z} $ .
Contrarily to this,  the flux appearing in the Dirac equation of the brane world--volume is always of type 
$ G_{a\bar b c \bar z} ~{\rm or} ~ G_{a \bar b z \bar c} ~. $
Thus for $N=2$ flux, the Dirac equation does not change at all and the zero 
modes are same as those of the fluxless case.

We now turn our attention to fluxes preserving $N=1$ supersymmetry. Such a flux is of the form
\beqa
G_4 \in \left(H^{2,0}(K3_1)\otimes H^{0,2}(K3_2)\right)
\oplus \left(H^{0,2}(K3_1)\otimes H^{2,0}(K3_2)\right) ~.
\label{neqoneflux}
\eeqa
In addition, it may contain flux of the form as given in Eq.(\ref{neq2flux}).
The susy conditions in presence of such a flux have been analyzed in great
detail in \cite{hep-th/0501139}. It has been realized there, that by an appropriate
choice of $(2,2)$ primitive flux, it is in fact possible to lift all the 
complex structure as well as K\"ahler moduli except the overall size of the 
$K3$. It has also been noticed that the fluxes of the type  given in 
eq.(\ref{neqoneflux}) stabilize both the $K3$s at an attractor point 
\cite{hep-th/0506014}. 
Attractive $K3$ surfaces are completely classified. They are in one-to-one 
correspondence with the (SL(2,Z) equivalent) matrices 
$$ Q = \left(\matrix{2a & b \cr b & 2c}\right) ,$$ 
where $a$, $b$ and $c$ are integers, and in addition $a$, $c$ and the the determinant 
of $Q$ are required to be positive. Two such matrices represent the same $K3$ 
if they are $SL(2,Z)$ equivalent. It has been shown in Ref.\cite{hep-th/0506014},
that the tadpole cancellation condition 
puts very strong constraints on the integers $a$,$b$ and $c$ appearing in the above 
matrix $Q$. Thus the $N=1$ solutions  are very limited and all of them
can be determined.

We now consider $M5$--branes wrapping divisors of the form $K3\times S$ in presence 
of such a flux. Locally, these fluxes are of the form $G_{ab\overline{cz}},
G_{a\overline{bc}z}$. The divisors under consideration have the cohomology 
\beqa
 H^{1,0}(K3\times P^1) = H^{3,0}(K3\times P^1) = 0 ~. 
\eeqa
Since $\phi_{\bar z}$ and $\phi_{\bar z\bar a\bar b}$ belong to these 
cohomology groups, they must be identically zero. We can now clearly see 
from the Dirac equations that the forms $\phi, \phi_{\overline{ab}}$ are 
harmonic, and in addition we have
\beqa
 g^{a\bar a'} g^{b\bar b'}  G_{ab\bar c\bar z} \phi_{\bar a'\bar b'} = 0 ~. 
\eeqa
This condition lifts the $\phi_{\bar a'\bar b'}$ mode. Hence, we only have 
massless modes corresponding to $\phi\in H^{0,0}(D)$. Note, that all the 
spinors also carry an $SO(2,1)$--index, and hence there is a doubling of 
massless modes. Since $H^{0,0}(D)$ is one--dimensional, we are now left with
two fermion zero modes, which is the right number for the instanton to 
contribute to $W_{np}$.

We now study the Dirac equations in presence of $(3,1)$-- and $(4,0)$--flux.
They take the simple form
\begin{eqnarray}
&& (\partial_{\bar c} \phi + 4 g^{b\bar b'} \partial_b \phi_{\bar b'\bar c}) 
 = 0\,, \cr &&  
 g^{a\bar a'} g^{b\bar b'}  G_{ab\bar c\bar z} \phi_{\bar a'\bar b'} = 0\,, \cr
&& \partial_{[\bar a} \phi_{\bar b\bar c]} = 0\,, \cr
&& \phi G_{\bar a\bar b\bar c\bar z} = 0\,.
\end{eqnarray}
Again, we find from the above that the forms $\phi,\phi_{\overline{ab}}$ are 
harmonic. In addition, we find that both zero modes $\phi$ as well as 
$\phi_{\overline{ab}}$ must be zero. Thus the presence of $(4,0)$--flux
lifts all the zero modes. As a result, we don't have any contribution to 
$W_{np}$ from the $M5$--instantons. 

We have seen in the above that we can choose an appropriate (2,2)
flux preserving 
$N=1$ susy, so that we have the correct number of fermion zero modes to have
a non-perturbative superpotential. But once we include 
a (4,0) flux, as enforced by the non-perturbative 
term in the susy conditions, all the zero modes are lifted which means 
that it is not consistent to keep the non-perturbative term. This raises 
a puzzle which we intend to resolve in the following section.

\section{Inclusion of the non-perturbative superpotential into the zero mode
counting}

In the last section, we have seen that a $(4,0)$--component of $G$ lifts all
zero modes. On the other hand,  the susy conditions
tell us that the $(4,0)$--part of $G$ is non-zero in the presence
of $W_{np}$. So there is an apparent mismatch.
The resolution of this puzzle seems to be to include $W_{np}$ into
the Dirac equation which determines the number of zero modes.
Then, $G_{4,0}$ should be balanced against $W_{np}$, as it is the case for
the susy conditions.

The Dirac part of the world--volume action on an $M5$--brane with fluxes has the form
\cite{hep-th/0503138}:
\begin{eqnarray}\label{M5action}
L_f^{M5}={1\over 2}\theta\lbrack \tilde\gamma^m\nabla_m+{1\over 24}
(\gamma^{\hat m\hat n\hat p}\tilde\gamma^qG_{q\hat m\hat n\hat p}
-\gamma^{\hat q}\tilde\gamma^{mnp}G_{mnp\hat q})\rbrack
\theta\, .
\label{diracmth}
\end{eqnarray}
For us, it is important to note that
the corresponding Dirac equation, whose solutions count the number
of fermionic zero modes, is
essentially determined
by the susy variation of the 11-dimensional gravitino field.
This can be seen as follows  \cite{hep-th/0507069}.
The supersymmetry conditions on the bulk, closed string background
are given by
\begin{eqnarray}
\delta\psi_M\epsilon=0\,,
\end{eqnarray}
which is the supersymmetry transformation of the 11-dimensional gravitino.
This can be translated to
the linear part 
of the Dirac equation from the world--volume action
as follows:
\begin{eqnarray}
(1-\Gamma_{M5})\Gamma^\alpha\delta \psi_\alpha\theta=0\, .
\label{fromspinor}
\end{eqnarray}
Here, $\delta\psi_\alpha$ is the pull--back of the gravitino variation to
the brane via $\delta\psi_\alpha=\delta\psi_M\partial_\alpha x^M$ and
 $\Gamma_\alpha=\Gamma_Ne^N_M\partial_\alpha x^M$. 
Therefore, one sees that the pull--back of the bulk gravitino equation is
equivalent to a solution of the Dirac equation.
Furthermore, one has to take into account the constraint from
$\kappa$-symmetry on the $M5$-brane:
\begin{eqnarray}
(1+\Gamma_{M5})\theta=0\, .
\end{eqnarray}
The number of zero modes is then given by the 
difference between the numbers of solutions of these two equations.

As we have already stated, we can recover the $M5$--brane world--volume action 
eq.(\ref{diracmth}) by using the explicit expressions for the internal gravitino 
variations in the absence of $W_{np}$ in eq.(\ref{fromspinor}).
We have already seen in \S2 that turning on $W_{np}$ alters the susy 
equations in the effective potential, 
as the effective superpotential now is $W=W_{flux}+W_{np}$. 
This addition should be described by the modulino equations, i.e.
$\delta\phi^i=\delta\phi^I=0$ should be now equivalent to $DW_{flux}+DW_{np}=0$.

 Substituting the 
expressions for the internal gravitino transformations with the general fluxes in ({\ref{M5action}}), one obatins
\begin{eqnarray}\label{diracset}
&& (\partial_{\bar c} \phi + 4 g^{b\bar b'} \partial_b \phi_{\bar b'\bar c}) 
- 2 g^{a\bar a'} g^{b\bar b'} g^{z\bar z}
 G_{ab\bar cz}  \phi_{\bar z\bar a'\bar b'} = 0\,,\cr
&& (\partial_{\bar a} \phi_{\bar z} 
+ 4 g^{b\bar b'} \partial_b \phi_{\bar z\bar a \bar b'} )  
+  g^{a\bar a'} g^{b\bar b'}  G_{ab\bar c\bar z} \phi_{\bar a'\bar b'}
= 0\,, \cr
&& \partial_{[\bar a} \phi_{\bar b\bar c]} 
+ {1\over 12} g^{z\bar z}\phi_{\bar z} G_{\bar a\bar b\bar c z} 
= 0\,, \cr
&& \partial_{[\bar a} \phi_{\bar z\bar b\bar c]} + {1\over 24} \phi
G_{\bar a\bar b\bar c\bar z}  = 0\, .
\end{eqnarray}

This is a set of local equations in the internal space. Every summand of  (\ref{diracset})  vanishes separately. 
This means that the set of equations
\begin{eqnarray}\label{cut}
G_{ab\ov c z}\phi^{abz}&=&0 \  ,\cr
G_{ab\ov {c z}}\phi^{ab}&=&0 \  ,\cr
G_{\ov {abc} z}\phi^{z}&=&0 \  ,\cr
G_{\ov {abc z}}\phi&=&0   
\end{eqnarray}
 is preventing the $\phi$, $\phi_{\ov a}$, $\phi_{\ov {ab}}$ and $\phi_{\ov {abz}}$ to be non-trivail zero-modes in the case of general flux $G_3$. 
On the other hand $G_{mnpq}$ correspond to the three-dimensional constant scalar fields which one obtains 
as coefficients by expansion of $G_3$ in the harmonic basis on $CY_4$:
\begin{eqnarray}
G_4 &=& G_{abcd} dz^a\wedge dz^b\wedge dz^c\wedge dz^d 
+ \sum_{i=1}^{h^{(3,1)}} G^i_{\overline{a}bcd} \omega^{i~\overline{a}bcd} \cr
&+&\sum_{k=1}^{h^{(2,2)}} G^k_{\overline{ab}cd}\tilde\omega^{k~\overline{ab}cd} 
+ \sum_{i=1}^{h^{(3,1)}} G^i_{\overline{abc}d} 
\bar\omega^{i~\overline{abc}d} \cr
&+& G_{\overline{abcd}}{ d\bar z}^{\bar a}\wedge{ d\bar z}^{\bar b}\wedge{ d\bar z}^{\bar c}\wedge{ d\bar z}^{\bar d}
\end{eqnarray}
with $\tilde\omega^k$ being basis elements of $H^{2,2}$. Since $H^{2,0} = 0$,
they can be expressed in terms of the basis elements $\omega^I$ of $H^{1,1}$ 
as $\tilde\omega^k = \sum_{I,J} \chi^k_{IJ} \omega^I\wedge\omega^J$.
The scalar fields $G$, $G^k$, $G^{i}$ are related to the flux superpotential by (\ref{Wflux1}) and (\ref{Wflux2}).
From the modulino equations (\ref{modulinomod}) we see that $W_{\rm flux}$ has to be replaced by $W=W_{\rm flux}+W_{\rm np}$.
This corresponds to the modification of $G$ to
\begin{eqnarray}
\hat G_{2,2} &:& \hat G_{ab\overline{cd}} = G_{ab\overline{cd}}\,, \cr
\hat G_{1,3}^i &:& \hat G_{a\overline{bcd}}^i = \epsilon_{a\overline{bcd}} D_i W=
G_{a\overline{bcd}}^i+\epsilon_{a\overline{bcd}}D_iW_{np}
\,, \cr
\hat G_{0,4} &:& \hat G_{\overline{abcd}} = \epsilon_{\overline{abcd}} {D_I W\over \del_I K}
=
G_{\overline{abcd}}+\epsilon_{\overline{abcd}}{D_I W_{np}\over \partial_IK}	
\,.
\end{eqnarray}

This amounts to modifying the world volume action (\ref{M5action}) in presenc of the
nonperturbative superpotential, where we now replace $G$ by $\hat G$. It is 
than straightforward to see that, using the susy conditions $D_iW = D_IW =0$,
the Dirac equation can be expressed as:
\begin{eqnarray}
&& (\partial_{\bar c} \phi + 4 g^{b\bar b'} \partial_b \phi_{\bar b'\bar c}) 
= 0\,, \cr
&& (\partial_{\bar a} \phi_{\bar z} 
+ 4 g^{b\bar b'} \partial_b \phi_{\bar z\bar a \bar b'} )  
+  g^{a\bar a'} g^{b\bar b'}  G_{ab\bar c\bar z} \phi_{\bar a'\bar b'}
= 0\,, \cr
&& \partial_{[\bar a} \phi_{\bar b\bar c]} = 0\,, \cr
&& \partial_{[\bar a} \phi_{\bar z\bar b\bar c]} = 0\,.
\end{eqnarray}
These conditions are identical to the ones coming from $(2,2)$ primitive flux
without $W_{np}$. The $(4,0)$-- and $(3,1)$--parts of the flux are compensated by 
the nonperturbative term. As a result, we find that the number of fermion zero
modes is unaltered. The apparent mismatch of the two answers in the previous
section was due to the fact that we had then ignored the back--reaction of 
the instanton on the background flux and the geometry. Once we take care of this
by modifying the fermionic terms accordingly, we obtain the expected result.

For the type $IIB$ Euclidean $D3$--brane, the story is very similar, hence
we will be very brief in the following. The Dirac Lagrangian can
be written in terms of the type $IIB$ gravitino variation, where in addition
also the dilatino variation appears:
\begin{eqnarray}
L_f^{D3}={1\over 2}e^{-\phi}\sqrt{\det g}~\bar\theta(1-\Gamma_{D3})
(\Gamma^\alpha\delta\psi_\alpha-\delta\lambda)\theta\, ,
\end{eqnarray}
where the bulk susy variations are $\delta\psi_m=0$ and
$\delta\lambda=0$. Substituting the expressions for  $\delta\psi_m$ and 
$\delta\lambda$ without $W_{np}$ into the above equation yields
\begin{eqnarray}
L^{D3} = 2 \sqrt{{\rm det}~g} ~~ \theta\left\{
e^{-\phi} \gamma^m \nabla_m + {1\over 8} \tilde G_{mn\hat p}
\gamma^{mn\hat p}\right\} \theta\, .
\end{eqnarray}
Once we use the modified expressions for $\delta\phi^k$, $\delta\phi^i$, $\delta\phi^I$ and $\delta\lambda$
in presence of $W_{np}$, we replace $G$ by 
\begin{eqnarray}
\hat G_{2,1} &:& \hat G_{ab\overline{c}} = \tilde G_{ab\overline{c}}\,, \cr
\hat G_{1,2}^i &:&  \hat G^i_{a\overline{bc}}=\tilde G^i_{a\overline{bc}} +\epsilon_{a\overline{bc}} D_i W_{np}
\,,\cr
\hat G_{0,3} &:& \hat G_{\overline{abc}} =\epsilon_{\overline{abc}} {D_IW \over \del_I K}= \tilde G_{\overline{abc}}+\epsilon_{\overline{abc}} {D_IW_{np}\over \del_I K}
\,, \cr
\hat G_{3,0} &:& \hat G_{\overline{abc}} = \epsilon_{abc} (\bar \tau - \tau) D_\tau W\,
= \tilde G_{\overline{abc}}+ \epsilon_{abc} (\bar\tau-\tau) D_I W_{np} \, .
\end{eqnarray} 

We can similarly analyze the Dirac equations. As expected, the number of 
fermion zero modes remains the same as in the case of primitive $(2,1)$--flux 
without the non-perturbative term.

\bigskip
\noindent {\large\bf Acknowledgments}\\[2ex]
We thank G. Cardoso, M. Haack, S. Mahapatra, S. Stieberger, S. Trivedi and M. Zagermann for useful discussion. 
S. Reffert and W. Schulgin thank the university of Munich for hospitality.

\appendix
\section{Dimensional reduction of $\delta\psi_m$}
We demonstrate the dimensional reduction of the supersymmetric variation of the gravitino on $CY_4$.

Firstly, we write the internal gravitino variation using holomorphic and antiholomorphic indices.
\begin{eqnarray}\label{var}
&&\delta \psi_e = \left[ \nabla_e + \frac{1}{24} \left(3 \gamma^{b\overline{cd}} 
G_{eb\overline{cd}} + \gamma^{\overline{bcd}} 
G_{ e\overline{bcd}}\right)\right]\xi \ , \cr
&&\delta \psi_{\bar e} = \left[ \nabla_{\bar e} + \frac{1}{24} \left(3 
\gamma^{b\overline{cd}} 
 G_{ \overline e b\overline{cd}} 
+ \gamma^{\overline{bcd}} 
G_{\overline{ebcd}}\right)\right]\xi \ .
\end{eqnarray}
$\psi_m$ is a vector-spinor, where $m$ is an internal vector index which transforms in the ${\bf 4\oplus\bar 4}$ representation of $SU(4)$. The spinor index of the eleven dimensional gravitino transforms in the ${\bf 32}$ under $SO(1,10)$. After compactification on a $CY_4$,  $SO(1,10)$  is broken to $SU(4)\times SO(2,1)$ and the spinor transforms in the ${\bf (1,2)\oplus (4,2) \oplus (6,2)\oplus  (\bar4,2)\oplus  (\bar1,2)}$. This means that  $\psi_e$ can be written as a sum of $(0,p)$-forms with one additional holomorphic or antiholomorphic index.
\begin{equation}\label{psi}
\psi_e=\phi_e|\Omega>+\phi_{e\bar a}\gamma^{\bar a}|\Omega>+\phi_{e\overline{ab}}\gamma^{\overline{ab}}|\Omega>+\phi_{e\overline{abc}}\gamma^{\overline{abc}}|\Omega>+\phi_{e\overline{abcd}}\gamma^{\overline{abcd}}|\Omega> \ .
\end{equation} 
Note that $\psi_e$ in (\ref{psi}) has an additional spinor index which transforms in the ${\bf 2}$ of $SO(1,2)$.
The rhs. of  (\ref{var}) is also such a spinor. $\xi$ can be written as $\xi=\epsilon\otimes \eta$, where $\eta$ is a covariantly constant spinor on the $CY_4$ and $\epsilon$ a supersymmetry parameter in the non-compact dimensions. We  write $\xi$ as
\begin{equation}
\xi=\hat\xi|\Omega>+\hat\xi_{\overline{abcd}}\gamma^{\overline{abcd}}|\Omega> \ . 
\end{equation}
and should remember that $\hat\xi$ has an additional index which transforms in the ${\bf 2}$ under $SO(1,2)$.
The rhs. of the first equation in (\ref{var}) is then
\begin{eqnarray}\label{expansion}
\delta \psi_e &=& \left[ \nabla_e + \frac{1}{24} \left(3 \gamma^{b\overline{cd}} 
G_{eb\overline{cd}} + \gamma^{\overline{bcd}} 
G_{ e\overline{bcd}}\right)\right]\xi  \cr
&=&\left[ \nabla_e + \frac{1}{24} \left(6 G_{eb\overline{cd}}g^{b\bar d}\gamma^{\bar c} 
 + G_{ e\overline{bcd}}\gamma^{\overline{bcd}} 
\right)\right]\xi  \cr
&=&\nabla_e\left(\hat\xi|\Omega>+\hat\xi_{\overline{abcd}}\gamma^{\overline{abcd}}|\Omega>\right)+ \frac{1}{4}G_{eb\overline{cd}}g^{b\bar d}\hat\xi\gamma ^{\bar c}|\Omega>+\frac{1}{24}G_{e\overline{bcd}}\hat\xi \gamma^{\overline{bcd}}|\Omega> \ .
\end{eqnarray}
The open index $e$ corresponds to a one-form index, which means that we have a collection of $(1,p)$-forms.\footnote{We can introduce a second set of gamma matrices, which will commute with the first one, so for example $\phi_{a_1\ldots a_p \bar a_1\ldots \bar a_q}\tilde\gamma^{a_1}\dots\tilde\gamma^{a_p}\gamma^{\bar a_1}\dots\gamma^{\bar a_q}|\Omega>$ will correspond to a $(p,q)$-form. Here we will omit the second set of gamma-matrices to make the equations more transparent. A detailed explanation of this formalism is given in Chapter 15 of \cite{Green:1987mn}.}
We compare the forms of the same type on both sides and obtain the following set of equations:
\begin{eqnarray}\label{set2}
\delta\left(\phi_{e\bar a}\gamma^{\overline a}|\Omega>\right)&=&\frac{1}{4}G_{eb\overline{cd}}g^{b\bar d}\hat\xi\gamma^{\bar c}|\Omega> \ , \cr
\delta\left(\phi_{e\overline{ab}}\gamma^{\overline{ab}}|\Omega>\right)&=&0 \ ,\cr
\delta\left(\phi_{e\overline{abc}}\gamma^{\overline{abc}}|\Omega>\right)&=& \frac{1}{24} G_{e\overline{bcd}}\hat\xi \gamma^{\overline{bcd}}|\Omega> \ .
\end{eqnarray}
These are the only forms from (\ref{expansion}), which do not vanish on a $CY_4$.

Let us look at the second equation of (\ref{var}) where the additional index is antiholomorphic. To see this index as a form index we have to make it holomorphic. This can be done by applying  Serre's generalization of Poincar\'e duality
\begin{equation}
\tilde\psi_{abc}=\psi_{\bar e}g^{e\bar e}\omega_{abce} \ ,
\end{equation}
where $\omega_{abce}$ is the $(4,0)$-form of the $CY_4$.
\begin{eqnarray}
\delta{\tilde\psi_{abc}}&=&g^{e\bar e}\omega_{abce}\left(\nabla_{\bar e}+\frac{1}{24}\left(3\gamma^{f\overline{gh}}G_{\overline{e}f\overline{gh}}+\gamma^{\overline{fgh}}G_{\overline{efgh}}\right)\right)\left(\hat\xi|\Omega>+\hat\xi_{\overline{ijkl}}\gamma^{\overline{ijkl}}\right)|\Omega>\cr
&=&g^{e\bar e}\omega_{abce}\left(\nabla_{\bar e}+\frac{1}{4}G_{\bar ef\overline{gh}}g^{ f\bar h}\gamma^{\bar g}+\frac{1}{24}G_{\overline{efgh}}\gamma^{\overline{fgh}}\right)\left(\hat\xi|\Omega>+\hat\xi_{\overline{ijkl}}\gamma^{\overline{ijkl}}\right)|\Omega> \ .
\end{eqnarray}
Again, comparing the forms of the same type gives us
\begin{eqnarray}\label{set1}
\delta\left(\tilde\phi_{abc\bar a}\gamma^{\bar a}\right)|\Omega>&=&\frac{1}{4}g^{e\bar e}\omega_{abce}G_{\overline{e}f\overline{gh}}g^{f\bar h}\gamma^{\bar g}\hat\xi|\Omega> \ , \cr
\delta\left(\tilde\phi_{abc\overline{ab}}\gamma^{\overline{ab}}\right)|\Omega>&=&0 \ , \cr
\delta\left(\tilde\phi_{abc\overline{abc}}\gamma^{\overline{abc}}\right)|\Omega>&=&\frac{1}{24}g^{e\bar e}\omega_{abce}G_{\overline{efgh}}\gamma^{\overline{fgh}}\hat\xi|\Omega> \ .
\end{eqnarray}

Eqs. (\ref{set2}) and (\ref{set1}) can be expanded in the basis of harmonic forms on the $CY_4$ as follows:
\begin{eqnarray}
\delta\left(\phi_i\omega^i_{(1,3)}\right) &=& g_i\omega^i_{(1,3)}\hat\xi \cr
\delta\left(\phi_I\omega^I_{(1,1)}\right) &=& g_I\omega^I_{(1,1)}\hat\xi \cr
\delta\left(\phi_i\omega^i_{(1,2)}\right) &=& 0 \cr
\delta\left(\phi_i\omega^i_{(2,3)}\right) &=& 0 
\end{eqnarray}
where $\omega^I_{(1,1)}$ and $\omega^i_{(1,3)}$ are basis elements of 
$H^{1,1}(CY_4)$  and $H^{1,3}(CY_4)$ respectively.

If we repeat the calculations for the type IIB case, we obtain an equation for the $(1,2)$-form, another one for the $(2,2)$-form and $(3,0)$-form for the dilatino:
\begin{eqnarray}\label{eq}
\delta\left(\phi_{e\overline{ab}}\gamma^{\overline{ab}}|\Omega>\right)& = &-\frac{1}{8}G_{e\overline{ab}}\gamma^{\overline{ab}}\hat\xi^*|\Omega>-\frac{1}{16}g^{a\overline c}g_{e\overline{a}}G_{a\overline{bc}} \gamma^{\overline{ab}}\hat\xi^*|\Omega>\ ,\cr
\delta\left(\phi_{\overline e |\overline{ab}}\gamma^{\overline{ab}}|\Omega>\right)& = &-\frac{1}{16}G_{\overline{abe}}\gamma^{\overline{ab}}\hat\xi^*|\Omega> \ , \cr
\delta\left(\lambda^*_{\overline{abc}}\gamma^{\overline{abc}}|\Omega>\right)&=&\frac{i}{4}\overline G_{\overline{abc}}\gamma^{\overline{abc}}\hat\xi^*|\Omega> \ .
\end{eqnarray}
The second equation corresponds to the $(2,2)$-form\footnote{Note, that in this notation the holomorphic indices correspond to the holomorphic part of the form and vice versa. The antiholomorphic index $\bar e$ has no meaning as form index before applying Serre's duality. That is why we put $|$ there to prevent  its mixing with the antiholomorphic indices.} after applying Serre's duality and to a $(1,1)$--form by forming the Hodge dual.

These equations (\ref{eq}) can be expanded in the basis of harmonic forms on the $CY_3$ and written then as 
\begin{eqnarray}
\delta\left(\phi_i \, \omega^i_{(1,2)}\right)&=&g_i\, \omega^i_{(1,2)}\hat\xi \ , \qquad\qquad i=1,\ldots ,h^{(2,1)} \ ,\cr
\delta\left(\phi_I\, \omega^I_{(1,1)}\right)&=&g_I\, \omega^I_{(1,1)}\hat\xi \ , \qquad\qquad I=1,\ldots ,h^{(1,1)} \ ,\cr
\delta\left(\lambda^{(0,3)}\omega_{(0,3)}\right)&=&g^{(0,3)}\omega_{(0,3)}\hat\xi \ .
\end{eqnarray}
$\phi_i$, $\phi_I$ and $\lambda^{(0,3)}$ correspond to the 4-dimensional complex structure modulinos, the K\"ahler modulinos and the dilatino respectively.

Finally, let us rewrite the variation of the modulino fields as it will be needed for our investigation:

\noindent
For the M-theory case:
\begin{eqnarray}
\delta\phi_{e\overline c}^k&=&\frac{1}{4}\left(G_{eb\overline{cd}}g^{b\overline d}\right)^k\hat\xi\ , \ \ \ \ \  \qquad\qquad k=1,\ldots, h^{(1,1)}  \ ,\cr
\delta\phi^i_{e\overline{abc}}&=&\frac{1}{24}G^i_{e\overline{abc}}\hat\xi \ , \qquad\qquad\qquad\qquad \  i=1,\ldots, h^{(3,1)} \ , \cr
\delta\phi^I_{\overline e | \overline{abc}}&=&\frac{1}{24}G_{\overline{eabc}}\hat\xi\ , \qquad\qquad\qquad\qquad I=1,\ldots, h^{(1,1)} \ .
\end{eqnarray}
For the type IIB case
\begin{eqnarray}
\delta \phi^i_{e\overline{ab}}&=&-\frac{1}{8}G_{e\overline{ab}}^i\hat \xi^*-\frac{1}{16}g^{a\overline c}g_{e\overline a}G^i_{a\overline{bc}}\hat\xi^* \ , \qquad i=1,\ldots ,h^{(2,1)} \ , \cr
\delta\phi^I_{\overline e |\overline{ab}}&=&-\frac{1}{16}G_{\overline{eab}}\hat\xi^* \ , \ \ \qquad\qquad\qquad\qquad \ I=1,\ldots , h^{(1,1)} \ , \cr
\delta\lambda^*_{\overline{abc}}&=&\frac{i}{4}\overline G_{\overline{abc}}\hat\xi^* \ .
\end{eqnarray}

We label the modulinos with the indices $k,i,I$. Additionally, they have indices from the beginning of the alphabet. 
Let us briefly comment about this.
 
 A $(p,q)$-form  $\nu$ can be expanded in the basis of harmonic $(p,q)$-forms $\omega^i$: $\nu=\nu_i\,\omega^i$. In the case of a complex manifold, the number of the harmonic forms is given by the corresponding Hodge number $h^{(p,q)}$. On the other hand we can write the form in every local patch as $\nu=\nu_{a_1\ldots a_p \ \overline{a}_1\ldots\overline{a}_q}dz^{a_1}\wedge\ldots\wedge dz^{a_p}\wedge dz^{\overline{a}_1}\wedge\ldots\wedge dz^{\overline{a}_q}$. If $\nu_{a_1\ldots a_p \ \overline{a}_1\ldots\overline{a}_q}$ are constant, they should correspond to the coefficients $\nu_i$.  The whole $\nu_{a_1\ldots a_p \ \overline{a}_1\ldots\overline{a}_q}$ in all coordinate patches span a vector space, in which so many $\nu_{a_1\ldots a_p \ \overline{a}_1\ldots\overline{a}_q}$ are linearly dependent by the transition functions  that the dimension of this vector space is $h^{(p,q)}$. The linearly independent combinations of $\nu_{a_1\ldots a_p \ \overline{a}_1\ldots\overline{a}_q}$ are then in one to one correspondence to the $\nu_i$.

\end{document}